\documentclass[10pt,prl,aps,twocolumn,showpacs,amsmath,amssymb]{revtex4-1}
\usepackage{graphicx}
\usepackage{subfigure}
\usepackage{ulem}
\usepackage{amsmath}
\usepackage{amssymb}
\usepackage{wrapfig}
\usepackage[utf8]{inputenc}  
\usepackage{color} 

\newcommand{\ket}[1]{\vert #1 \rangle}

\begin{document}

\title{Control of spatial correlations between Rydberg excitations using rotary echo}
\author{N.~Thaicharoen*}
\author{A.~Schwarzkopf$^{\dag}$}
\author{G.~Raithel}
\affiliation{Department of Physics, University of Michigan, Ann Arbor, Michigan 48109, USA}
\date{\today}

\begin{abstract}
We manipulate correlations between Rydberg excitations in cold atom samples using a rotary-echo technique. The correlations are due to interactions between the Rydberg atoms.
In the rotary-echo excitation sequence, the phase of the excitation pulse is flipped at a selected time during the pulse. We measure the resultant change in the spatial pair correlation function of the excitations via direct position-sensitive atom imaging. For zero detuning of the lasers from the interaction-free Rydberg-excitation resonance, the pair-correlation value at the most likely nearest-neighbor Rydberg-atom distance is substantially enhanced when the phase is flipped at the middle of the excitation pulse.  In this case, the rotary echo eliminates most uncorrelated (un-paired) atoms, leaving an abundance of correlated atom pairs at the end of the sequence. In off-resonant cases, a complementary behavior is observed. We further characterize the effect of the rotary-echo excitation sequence on the excitation-number statistics of the atom sample.
\end{abstract}


\maketitle

Recently there has been a growing interest in applications of correlated Rydberg-atom systems. The correlations arise from strong electrostatic interactions between the Rydberg atoms, which lead to an excitation blockade~\citep{lukin_dipole_2001, urban_observation_2009, weber_mesoscopic_2015}. The correlations have been utilized to realize non-classical light sources, such as single-photon sources as well as sources of correlated photons \citep{dudin_strongly_2012}. Blockaded  Rydberg-atom samples can exhibit sub-Poissonian statistics for the Rydberg-atom number distribution \citep{liebisch_atom_2005, malossi_full_2014,schempp_full_2014}. Higher-order correlation leads to Rydberg-atom aggregation\citep{amthor_evidence_2010, garttner_dynamic_2013, schempp_full_2014, lesanovsky_out--equilibrium_2014, urvoy_strongly_2015} and crystallization of Rydberg excitations~\citep{pohl_dynamical_2010, schaus_observation_2012, schaus_crystallization_2015}.

In previous studies, the correlations between Rydberg atoms have been enhanced by detuning the excitation laser to match the van der Waals energy level shift at a well-defined pair separation~\citep{schwarzkopf_spatial_2013, thaicharoen_measurement_2015}. Another method to control Rydberg-atom correlations, proposed by W{\"u}ster \textsl{et. al.}~\citep{wuster_correlations_2010}, is based on a rotary excitation echo. As found previously, in simulations~\citep{hernandez_simulations_2008} and in experiments~\citep{raitzsch_echo_2008,younge_rotary_2009}, a rotary Rydberg-atom excitation echo occurs when the sign of the Rabi frequency is flipped in the middle of the excitation pulse. The rotary-echo technique can also be used in Rydberg-atom interferometry \citep{dunning_composite_2014, wang_atom-interferometric_2015}. Of particular relevance to this paper, it was predicted in~\citep{wuster_correlations_2010} that a rotary-echo excitation pulse leads to an enhancement of the Rydberg pair correlation function at a separation near the blockade radius.

\begin{figure}[b] 
\centering
\includegraphics[width=\linewidth]{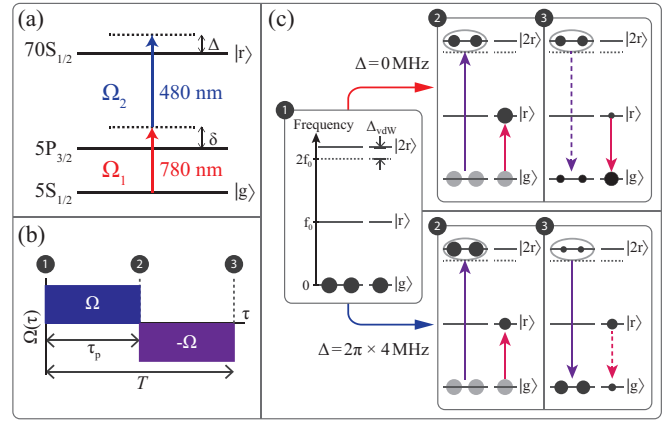}
\caption{(a) Two-photon excitation of a single $^{85}$Rb Rydberg atom. The detuning from the intermediate state $5P_{3/2}$ is $\delta=2 \pi \times 131$~MHz. The detuning $\Delta$ from the Rydberg $70S_{1/2}$ level is 0 or $2\pi \times 4$~MHz; in the latter case the detuning compensates the van der Waals interaction between the atoms in the multi-atom excitation of panel (c) (which is positive in our case). (b) Echo sequence during excitation. We apply an excitation pulse of duration $T=250$~ns. The two-photon Rabi frequency is switched from $\Omega$ to $-\Omega$ (in the field picture) at time $\tau_{\rm{p}}$. The time labels 1, 2, and 3 above the timing sequence correspond with the diagrams in (c). (c) Internal dynamics of atoms during the echo sequence. The frequency to excite one Rydberg atom $\ket{r}$ without interactions is $f_{0}$. The frequency leading to direct excitation of two Rydberg atoms (state $\ket{2r}$) is $f_{0}+\Delta_{\rm{vdW}}/2$. The circle sizes in the level diagrams illustrate the populations of van-der-Waals-interacting atom pairs (left pair of circles) and of isolated atoms (right circle) at times 1, 2 and 3 for the cases $\Delta=0$ and $\Delta = 2\pi \times 4$~MHz.}
\label{fig:timing}
\end{figure}

In the present work, we employ a rotary-echo sequence to excite cold $^{85}$Rb atoms into Rydberg states. We use a position-sensitive Rydberg-atom imaging and counting method~\citep{schwarzkopf_spatial_2013} to explore the effects of the echo on spatial Rydberg-atom correlations and on the excitation-number statistics. Both these observables for correlation in cold Rydberg-atom ensembles have been studied previously in the absence of a rotary echo (see~\citep{schwarzkopf_imaging_2011, schwarzkopf_spatial_2013}  and~\citep{liebisch_atom_2005, malossi_full_2014,schempp_full_2014}, respectively). Here, we  demonstrate that a rotary-echo sequence can substantially enhance spatial correlations between the Rydberg excitations prepared within the atom samples. We further observe that the rotary echo strongly affects the counting number statistics of Rydberg excitations in the samples. Studying cases of different laser detunings relative to the interaction-free atomic transition, we  establish that the rotary-echo-induced features strongly depend on the laser detuning. Particularly, it is seen that the rotary echo gives rise to strong spatial correlations when the excitation lasers are on-resonant, while in certain off-resonant cases the echo causes the opposite effect, namely the destruction of spatial correlations that would otherwise be present.

\begin{figure}[t] 
\centering
\includegraphics[width=\linewidth]{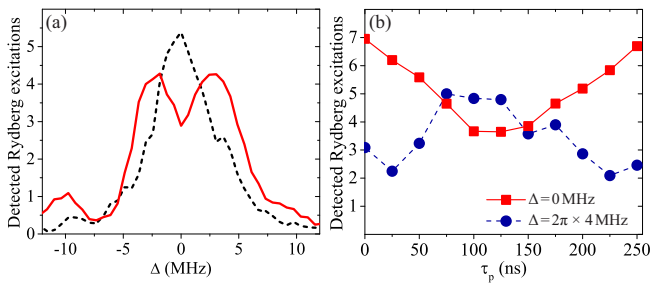}
\caption{(a) Rydberg excitation spectra for $\tau_{\rm{p}}=0$ (black dashed line) and 125~ns (red solid line). (b) Number of detected Rydberg excitations as a function of phase flip time $\tau_{\rm{p}}$ for $\Delta=0$~MHz (red squares) and $\Delta=2 \pi \times 4$~MHz (blue circles).}
\label{fig:spectra_counts}
\end{figure}

Schematics of the experiment and the timing sequence are presented in Fig.~\ref{fig:timing}. Cold $^{85}$Rb atoms in the $5S_{1/2}$ state are excited to the $70S_{1/2}$ state via two-photon excitation using 780- and 480-nm laser pulses that overlap in time and have a duration $T=250$~ns. The 780-nm laser has a Gaussian beam parameter w$_{0}=0.75$~mm and  a power of 600~$\mu$W. The 480-nm laser has a w$_{0}\approx 8~\mu$m and a power of 30 mW. The detuning from the intermediate state 5P$_{3/2}$ is $\delta= 2 \pi \times 131$~MHz. At the beam center, the Rabi frequencies of the lower and upper transitions are $\Omega_{1}=2\pi\times20$~MHz and $\Omega_{2}=2\pi\times21$~MHz, respectively, leading to a two-photon (one 780~nm and one 480~nm photon) Rabi frequency for single-atom excitation at the two-photon resonance ($\Delta$ = 0) of $\Omega=\Omega_{1}\Omega_{2}/(2\delta)=2\pi\times1.67$~MHz. In the case of substantial detuning $\Delta > 0$, Rydberg atoms may instead be directly excited in pairs at a separation where the detuning $\Delta$ matches the van der Waals interaction (see next paragraph). When $\Delta = 2 \pi \times 4~$MHz, as used in some of our experiments, the Rabi frequency for the simultaneous excitation of such pairs is $\Omega_{\rm{pair}}=\Omega^2/(2\Delta)=2\pi \times 0.35$~MHz. 

At $\Delta = 2 \pi \times 4~$MHz, atom pairs are efficiently excited at a measured pair separation $R \approx 10~\mu$m, where the excess energy associated with the laser excitation of the atoms equals the van-der-Waals energy shift, $\Delta \approx C_6/R^6$, where $C_6$ is the van-der-Waals dispersion coefficient~\citep{schwarzkopf_spatial_2013, thaicharoen_measurement_2015}. This effect arises because the presence of Rydberg atoms in the sample facilitates the addition of new Rydberg atoms at a fairly well-defined facilitation radius that depends on laser detuning and interaction strength~\citep{amthor_evidence_2010, garttner_dynamic_2013, schempp_full_2014, lesanovsky_out--equilibrium_2014, urvoy_strongly_2015}. In the present work, we employ this effect to prepare an abundance of Rydberg-atom pairs at a well-defined internuclear separation.

To implement a rotary echo, we shift the phase of the radio frequency (RF) signal that is applied to the acousto-optic modulator that determines the optical phase of the 480-nm laser pulse. We use a $180^\circ$-power splitter together with a high-isolation RF switch to change the phase of the RF signal by $\pi$ at time $\tau_{\rm{p}}$ after the excitation lasers are turned on, as shown in Fig.~\ref{fig:timing}(b). The phase-flip time $\tau_{\rm{p}}$ is varied from 0 to 250~ns, with a step size of 25~ns. Immediately after the excitation pulse, we field-ionize the Rydberg atoms by applying a high voltage to a tip imaging probe (TIP). The field ions are counted and their positions are measured using a position-sensitive microchannel plate and phosphor screen assembly (MCP) with a CCD camera. Each CCD image contains ion blips whose positions represent the 2-dimensional projection of the positions of the parent Rydberg atoms in the excitation region. The excitation region is determined by the position of the 480-nm beam, which is centered 470~$\mu$m above the TIP. At this distance, the ion-imaging setup provides a magnification of 150 (see Ref.~\citep{thaicharoen_measurement_2015} for the magnification calibration process). For each detuning $\Delta$ and flip time $\tau_p$ we take 10000 images and choose the 5000 images with the highest numbers of excitations to calculate  pair correlation images. Angular integrals of the pair correlation images yield the radial pair correlation functions $I(R)$. Quantitative information on the echo effect is then extracted from the $I(R)$.

\begin{figure*}[t!] 
\centering
\includegraphics[width=\textwidth]{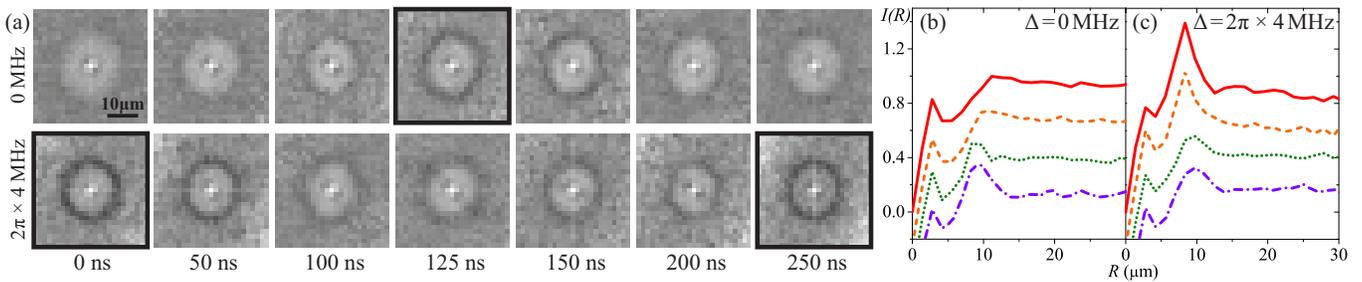}
\caption{(a) Pair correlation images at selected $\tau_{\rm{p}}$ for (top row) $\Delta=0$~MHz and (bottom row) $\Delta=2 \pi \times 4$~MHz. The linear grayscale ranges from 0 (white) to 2 (black) where values of 1, $<$1, and $>$1 indicate no correlation, anticorrelation, and positive correlation, respectively.  The bold borders indicate pair correlation images obtained from systems that contain mostly pair-excitations. The 3 to 4 $\mu$m pattern near the center of each image is an experimental artifact due to ion feedback (see text). (b) Angular integrals $I(R)$ of the pair correlation images in (a) for $\Delta=0$~MHz for $\tau_{\rm{p}}=$ 0, 50, 100, and 125~ns (top to bottom). The vertical axis is for the 0~ns curve. The other curves are shifted down in equidistant intervals of 0.3 for clarity. (c) Same as (b), but for $\Delta=2 \pi \times 4$~MHz.}
\label{fig:paircorr_all}
\end{figure*}

To verify that a rotary echo~\citep{hernandez_simulations_2008, raitzsch_echo_2008,younge_rotary_2009} occurs in our experiment, we measure Rydberg excitation spectra for $\tau_{\rm{p}}=0$ and 125~ns, as shown in Fig.~\ref{fig:spectra_counts}(a). The peak position in the echo-free spectrum ($\tau_{\rm{p}} = 0$) marks the on-resonant transition $\Delta=0$. The echo occurs when $\tau_{\rm{p}}$ is close to half of the pulse duration. In this case, the spectra are broader and have a depression of detected Rydberg counts at $\Delta = 0$ [Fig.~\ref{fig:spectra_counts}(a)]; the $\Delta = 0$ counts have a minimum when $\tau_{\rm{p}}=T/2=125$~ns [Fig.~\ref{fig:spectra_counts}(b)], as expected for a rotary echo. 

In Fig.~\ref{fig:spectra_counts}(a), the signal does not drop to zero at $\Delta = 0$, as would be the case for a perfect echo, due to the following reasons. First, slightly off-resonantly excited atom pairs whose pair energies are shifted due to the van der Waals interactions do not undergo a perfect echo; this is the effect we essentially exploit later in this work. Second, the spectrum is convolved with the profile of the shot-to-shot laser-frequency jitters. We have simulated the excitation spectra of atom pairs in  a disordered atomic sample and convolved the results with a Gaussian profile for the frequency jitter. Assuming a full-width-at-half-maximum (FWHM) of 3 to 4~MHz for the Gaussian profile, we obtain simulated spectra that are consistent with the experimentally observed ones. Of lesser importance is the residual Doppler effect, which may broaden spectral structures by several 100~kHz. We further believe that phase noise of the lasers during individual excitation sequences is not important.

We now turn to describing the effect of the echo sequence on the spatial pair-correlation functions and the Rydberg-atom-counting statistics. The pair correlation images measured for $\Delta = 0$~MHz are shown in the top row of Fig.~\ref{fig:paircorr_all}(a). When $\tau_{\rm{p}}$ is increased from 0 or decreased from $T$ towards $T/2$, the pair correlation image clearly develops an enhancement ring with a radius of about $10~\mu$m. Visual inspection of the images in Fig.~\ref{fig:paircorr_all}(a) shows that the enhancement ring reaches maximal  contrast for $\tau_{\rm{p}}=T/2=125$~ns. For the case $\Delta=2 \pi \times 4$~MHz (bottom row of Fig.~\ref{fig:paircorr_all}(a)), the pair correlation is maximally enhanced at a radius near 10~$\mu$m when $\tau_{\rm{p}}=0$ or $\tau_{\rm{p}}=T$, while it becomes washed out when $\tau_{\rm{p}}$ approaches $T/2$. A pair-correlation signal at very short distances (about 3~$\mu$m) is an artifact caused by ion feedback from the TIP; field electrons impacting on the TIP release a secondary ion that impacts the MCP close to the primary (Rb$^+$) ion~\citep{thaicharoen_atom-pair_2016}. In our quantitative analysis below, we employ the angular integrals $I(R)$ of the pair correlation images. We show the $I(R)$ curves in Fig.~\ref{fig:paircorr_all}(b) for the case $\Delta = 0$, and in Fig.~\ref{fig:paircorr_all}(c) for the case $\Delta=2 \pi \times 4$~MHz. 

In the explanation of the pair correlation data and the strength of the correlation enhancement, we concentrate on the case $\tau_{\rm{p}} = T/2$ and refer to Fig.~\ref{fig:timing}. The two-photon Rabi frequency ($\Omega=2\pi\times1.67$~MHz) is high enough that many excitation domains within the sample carry one Rydberg excitation (state $\ket{r}$) after the first phase of the excitation pulse ({\sl{i.e.}}, at time $\tau_{\rm{p}}$; isolated atom in Fig.~1(c)). However, due to the bandwidth of the excitation pulse, some excitation domains become populated with a Rydberg-excitation pair (state $\ket{2r}$) via a two-photon process. The doubly-excited domains have an energy level shift $\Delta_{\rm{vdW}}$ from the drive field due to the van der Waals interaction (atom pairs in Fig.~\ref{fig:timing}(c)). The atom populations after flipping the phase and completing the excitation pulse are illustrated in Fig.~\ref{fig:timing}(c), at time labels 3. For the case $\Delta = 0$, domains in the state $\ket{r}$ are de-excited back to the $\ket{g}$-state. However, the domains in excited state $\ket{2r}$ are off-resonant, leaving them with some probability in $\ket{2r}$ after completion of the echo sequence. In effect, after the sequence we expect to find a relative over-abundance of paired atoms, at a correlation distance slightly smaller than the blockade radius. The experimental data support this scenario. The opposite behavior is observed in the case $\Delta = 2\pi \times 4$~MHz. In that case, domains in the state $\ket{2r}$ undergo an echo (for $\tau_{\rm{p}}=T/2$) and are de-excited back to the $\ket{g}$-state. This leaves an abundance of isolated Rydberg atoms in the state $\ket{r}$ after the echo sequence, with a reduced degree of correlation. 

\begin{figure}[t] 
\centering
\includegraphics[width=\linewidth]{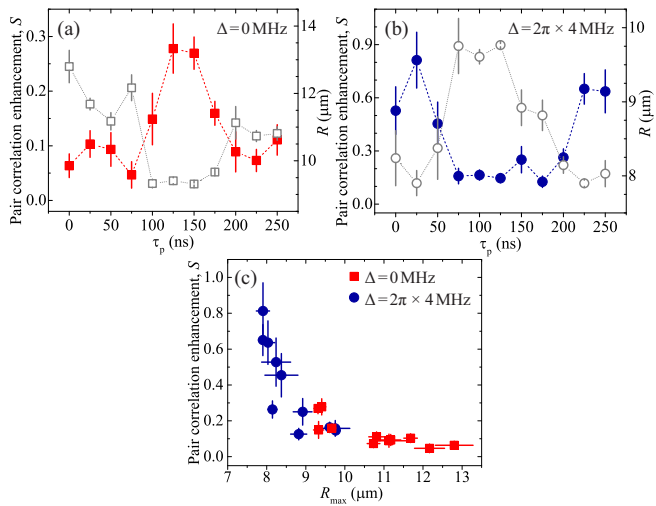}
\caption{Pair correlation enhancement, $S$ (red squares and blue circles) and $R_{\rm{max}}$ (gray hollow squares and gray hollow circles) as a function of $\tau_{\rm{p}}$ for (a) $\Delta=0$~MHz and (b) $\Delta=2 \pi \times 4$~MHz, respectively. (c) Pair correlation enhancement $S$ versus $R_{\rm{max}}$ for both cases (a) and (b).}
\label{fig:visibility}
\end{figure}

For a more quantitative analysis of the data, from the $I(R)$ curves we obtain peak values $I_{\rm{max}}$ and radii $R_{\rm{max}}$ of maximal correlation enhancement, determined by the local parabolic fits to the maxima near $R = 10~\mu$m. We calculate $S$, the strength of the correlation enhancement of a pair correlation function, as:
\begin{equation}
S = (I_{\rm{max}} - \langle I \rangle)/\langle I \rangle\quad.
\end{equation}
We determine the asymptotic values $\langle I \rangle$ by taking the average of $I(R)$ over the range 20~$\mu$m $< R <60~\mu$m. 
Figure~\ref{fig:visibility} shows the dependence of the pair enhancements $S$ and the radii $R_{\rm{max}}$ on $\tau_{\rm{p}}$; panels (a) and (b) show the cases $\Delta = 0$~MHz and $\Delta = 2 \pi \times 4$~MHz, respectively.
As anticipated from Fig.~\ref{fig:paircorr_all}, we see in Fig.~\ref{fig:visibility}(a) and Fig.~\ref{fig:visibility}(b) that the curves $S(\tau_{\rm{p}})$ exhibit pronounced maxima at $\tau_{\rm{p}}=T/2$, in the case $\Delta = 0$, and at $\tau_{\rm{p}} = 0$ and $T$, in the case $\Delta = 2\pi \times 4$~MHz. Thus, the curves $S(\tau_{\rm{p}})$ for the cases $\Delta = 0$ and $2\pi \times 4$~MHz exhibit opposite trends. Close inspection of the $I(R)$ curves in Figs.~\ref{fig:paircorr_all}(b-c) also shows that the FWHM of the enhancement peak decreases when the $I_{\rm{max}}$ increases.

Figures~\ref{fig:visibility}(a) and Fig.~\ref{fig:visibility}(b) reveal an additional subtle trend. In both the $\Delta = 0$~MHz and the $\Delta = 2 \pi \times 4$~MHz cases, the values for $R_{\rm{max}}$ are smaller for larger values of $S$. The relationship between $R_{\rm{max}}$ and $S$ is shown in Fig.~\ref{fig:visibility}(c). It is clearly seen that the pair correlation enhancement $S$ increases from near zero to $\sim 0.8$ as the radius of maximal correlation diminishes from $\sim 13~\mu$m to $\sim 8 ~\mu$m. The underlying reason for this behavior is that large correlation enhancement $S$ corresponds to strong van der Waals interaction, $C_{6}/R^{6}$; hence a large $S$ corresponds to a small $R$, regardless of the exact excitation conditions. This physical connection is explicitly expressed in Fig.~\ref{fig:visibility}(c). It is noteworthy that the $S(R_{\rm{max}})$ dependencies extracted from the $\Delta = 0$ and the $\Delta = 2 \pi \times 4$~MHz measurements agree very well in Fig.~\ref{fig:visibility}(c) within their overlap region $9~\mu$m $<R_{\rm{max}}<10~\mu$m.

\begin{figure}[t] 
\centering
\includegraphics[width=0.94\linewidth]{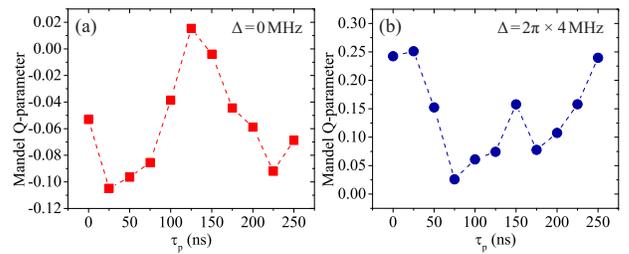}
\caption{Mandel Q-parameter as a function of $\tau_{\rm{p}}$ for (a) $\Delta=0$~MHz and (b) $\Delta=2 \pi \times 4$~MHz.}
\label{fig:Qvalue}
\end{figure}

We have established that the echo excitation sequence has a profound effect on spatial correlations in cold Rydberg-atom samples. We expect a complementary effect in the Rydberg excitation counting statistics, which should also carry signatures of the correlations in the system. A common measure to characterize counting-statistics is the Mandel $Q$-parameter~\citep{liebisch_atom_2005, ates_strong_2006, schempp_full_2014, malossi_full_2014}, which is calculated from
\begin{equation}
Q=\frac{\left<n(\tau_{\rm{p}})^{2}\right>-\left<n(\tau_{\rm{p}})\right>^{2}}{\left<n(\tau_{\rm{p}})\right>}-1\quad.
\label{eq:qvalue}
\end{equation}
Figure~\ref{fig:Qvalue} shows the $Q$-parameters as a function of $\tau_{\rm{p}}$. The plotted values do not take the detection efficiency $\eta$ into account; the actual $Q$-parameters in the sample are given by $Q/\eta$. The Q-parameters observed for $\Delta=0$~MHz are mostly negative, ranging from  -0.10 to 0.02, while the ones for $\Delta=2 \pi \times 4$~MHz are positive, ranging from 0.03 to 0.25. This shows that for $\Delta=0$ and $\tau_{\rm{p}}\sim 0$ and $\sim T$ the atoms largely follow sub-Poissonian statistics ($Q<0$), while for $\Delta=2\pi \times 4$~MHz and $\tau_{\rm{p}}=0$ and $T$ they follow super-Poissonian statistics ($Q>0$). 

In the following we explain the trends observed in our Q-value measurements. The Q-parameters for systems containing mostly pair-excitations are expected to trend toward positive values~\citep{wuster_correlations_2010, heeg_hybrid_2012}, indicating super-Poissonian distributions.  This effect is demonstrated in Fig.~\ref{fig:Qvalue}, in which the Q-parameters reach high values when pair excitations are dominant. This is the case for $\tau_{\rm{p}}=125$~ns at $\Delta=0$~MHz, and $\tau_{\rm{p}}=0$ and 250~ns at $\Delta=2 \pi \times 4$~MHz. These cases correspond to the images with bold borders in Fig.~\ref{fig:paircorr_all} (a).

For a qualitative explanation of why ensembles of correlated Rydberg-atom pairs lead to super-Poissonian statistics, we consider a Poissonian distribution of atom pairs ($Q_{\rm{pair}}=0$). Since every pair contains two atoms, we substitute $n=2n_{\rm{pair}}$ into Eq.~\ref{eq:qvalue}. The relation between $Q_{\rm{pair}}$ and the Q-value for single-atom detections, $Q_{\rm{single}}$, is seen to be
\begin{equation}
Q=Q_{\rm single}=2Q_{\rm pair}+1\quad.
\end{equation}

Hence, a Poissonian distribution of uncorrelated Rydberg-atom pairs ($Q_{\rm pair}=0$) results in a super-Poissonian distribution of single-atom detections ($Q_{\rm single}=1$). Considering that the detection efficiency is $\eta \sim 0.3$, we expect a measured $Q \sim 0.3$, which is close to the value shown in Fig.~\ref{fig:Qvalue}(b) at $\tau_{\rm{p}}=0$ and $T$.

In summary we have measured the effect of the rotary echo on spatial pair correlation functions and counting statistics of Rydberg atoms for on- and off-resonant excitations. The measurements exhibit a fundamental connection between spatial correlations and counting statistics. The result also shows that it is possible to prepare correlated Rydberg atoms at a well-defined interatomic separation by using this method. This ability could be useful in atom kinetics experiments~\citep{thaicharoen_measurement_2015, thaicharoen_atom-pair_2016}. We believe that the pair correlation enhancement  afforded by the echo can be improved, in the future, by reducing the excitation bandwidth. Further, it would be interesting to explore the effect of the echo sequence in the case of anisotropic interaction potentials.

This work was supported by the NSF (PHY-1205559 and PHY-1506093) and the AFOSR (FA9550-10-1-0453).  N.T. acknowledges support from DPST of Thailand. We thank S. W\"uster for helpful discussions.\\

\noindent*nithi@umich.edu\\
$^{\dag}$Present address: zeroK NanoTech Corporation, Gaithersburg, MD 20879, USA

\bibstyle{apsrev4-1}
%

\end{document}